\documentclass[twocolumn,prb]{revtex4}
\usepackage{amsfonts}
\usepackage[T1]{fontenc}
\usepackage{amsmath,amsbsy,amssymb,graphicx}
\usepackage{times}

\let\mathbf=\boldsymbol

\begin{document}

\title{Non-Hermitian higher-order topological states in nonreciprocal and
reciprocal systems\\
with their electric-circuit realization }
\author{Motohiko Ezawa}
\affiliation{Department of Applied Physics, University of Tokyo, Hongo 7-3-1, 113-8656,
Japan}

\begin{abstract}
A prominent feature of some one-dimensional non-Hermitian systems is that 
all right-eigenstates of the non-Hermitian Hamiltonian are localized
in one end of the chain.  
The topological and trivial phases are
distinguished by the emergence of zero-energy modes within the skin states 
in the presence of the chiral symmetry. Skin states are formed when the system is nonreciprocal, 
where it is said nonreciprocal if the absolute values of the right- and
left-going hoppings amplitudes are different. Indeed, the zero-energy edge
modes emerge at both edges in the topological phase of the reciprocal
non-Hermitian system. Then, analyzing higher-order topological insulators in
nonreciprocal systems, we find the emergence of topological zero-energy
modes within the skin states formed in the vicinity of one corner.
Explicitly we explore the anisotropic honeycomb model in two dimensions and
the diamond lattice model in three dimensions. We also study an
electric-circuit realization of these systems. Electrical circuits with
(without) diodes realize the nonreciprocal (reciprocal) non-Hermitian
topological systems. Topological phase transitions are observable by
measuring the impedance resonance due to zero-admittance topological corner
modes.
\end{abstract}

\maketitle

\textit{Introduction: }Topological physics is one of the most important
achievements in contemporary physics, among which there are topological
insulators and its generalization to higher-order topological insulators\cite%
{Fan,Science,APS,Peng,Lang,Song,Bena,Schin,EzawaKagome,Bis,Khalaf, Switch}.
They are characterized by the bulk topological numbers, where the
bulk-boundary correspondence and its generalization play a key role. In
particular, topological zero-energy corner modes emerge for the second-order
topological insulators in two dimensions and for the third-order topological
insulators in three dimensions. They have been studied not only in condensed
matter physics but also in various systems such as photonic\cite%
{Photonics,Gold,Photon}, phononic\cite{Suss,Aco,Phonon,Xue,Alex} and
microwave\cite{Hu,Microwave} systems. $LC$ electric circuits have also
topological phases\cite{ComPhys,TECNature,Garcia,Hel,Lu,EzawaTEC}.

Recently, non-Hermitian topological systems attract increasing attentions%
\cite%
{Bender,Bender2,Konotop,Gana,Lee,Rako,Lieu,Yin,Yao,Jin,Liang,Nori,Fu,Menke}.
They are realized in photonic systems\cite{Mark,Scho,Pan,Weimann}, microwave
resonators\cite{Poli}, wave guides\cite{Zeu}, quantum walks\cite{Rud,Xiao}
and cavity systems\cite{Hoda}. Non-Hermitian generalizations of the
Su-Schrieffer-Heeger (SSH) model have been most studied\cite%
{Scho,Zhu,Weimann,Poli,Lee,Lieu,Yin,Yao,Kunst}. Non-Hermitian systems have
new aspects. First, we must differentiate between the right and left
eigenenergies and eigenstates, which are defined by $H\left\vert \psi ^{%
\text{R}}\right\rangle =\varepsilon ^{\text{R}}\left\vert \psi ^{\text{R}%
}\right\rangle $ and $H^{\dagger }\left\vert \psi ^{\text{L}}\right\rangle
=\varepsilon ^{\text{L}}\left\vert \psi ^{\text{L}}\right\rangle $. The
right and left eigenenergies are complex in general. A prominent property of
some non-Hermitian systems is the non-Hermitian skin effect, where 
all right-eigenstates are localized
in one end of a finite chain with the bulk spectrum being totally
modified\cite{Xiong,Mart,UedaPRX,Yao,Kunst,Jin}. An interesting feature is
that the topological phase transition point for a finite system is different
from the bulk gap closing point. The topological and trivial phases are
distinguished by the presence of the zero-energy edge mode within the skin
states. The topological number is given by the so-called non-Bloch
topological invariant\cite{Yao,Yao2,Yang}, which well describes the phase
transition point for a finite system.

\begin{figure*}[t]
\centerline{\includegraphics[width=0.88\textwidth]{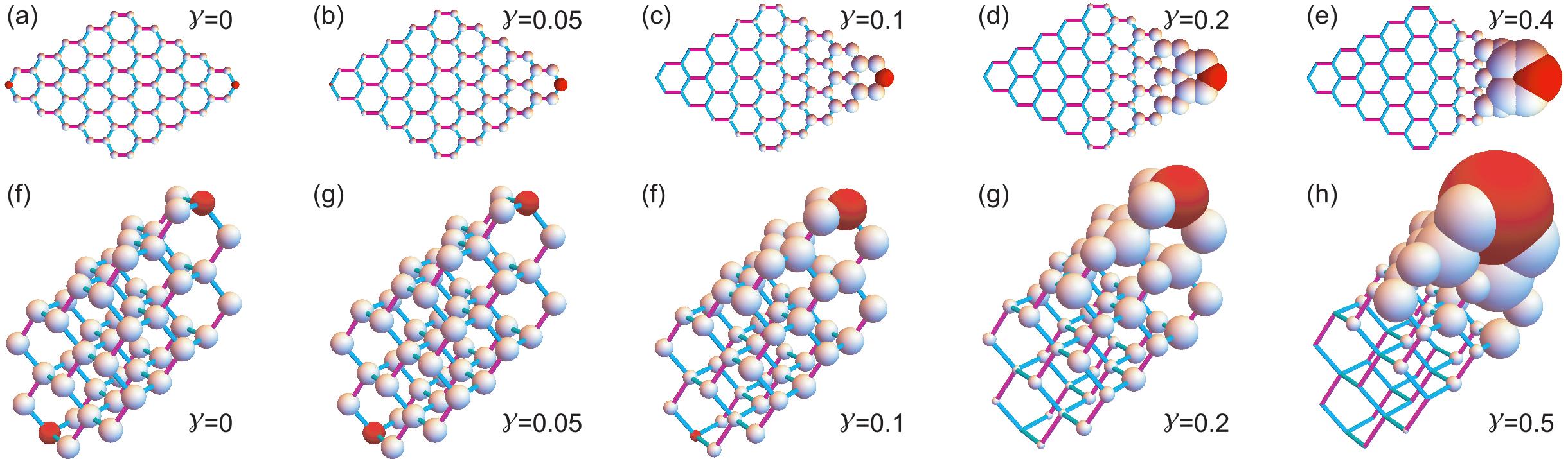}}
\caption{ (a)--(e) Development of the skin states (in gold) and the
topological corner modes (in red) in rhombus geometry of the nonreciprocal
honeycomb system as the nonreciprocity $\protect\gamma$ increases. The size
of a ball represents the magnitude of LDOS. We have set $t_{A}=0.4$ and $%
t_{B}=1$. (f)--(h) The corresponding ones in rhombohedron geometry of the
diamond lattice system. We have set $t_{A}=0.5$ and $t_{B}=1$. }
\label{FIG1}
\end{figure*}

\begin{figure}[t]
\centerline{\includegraphics[width=0.49\textwidth]{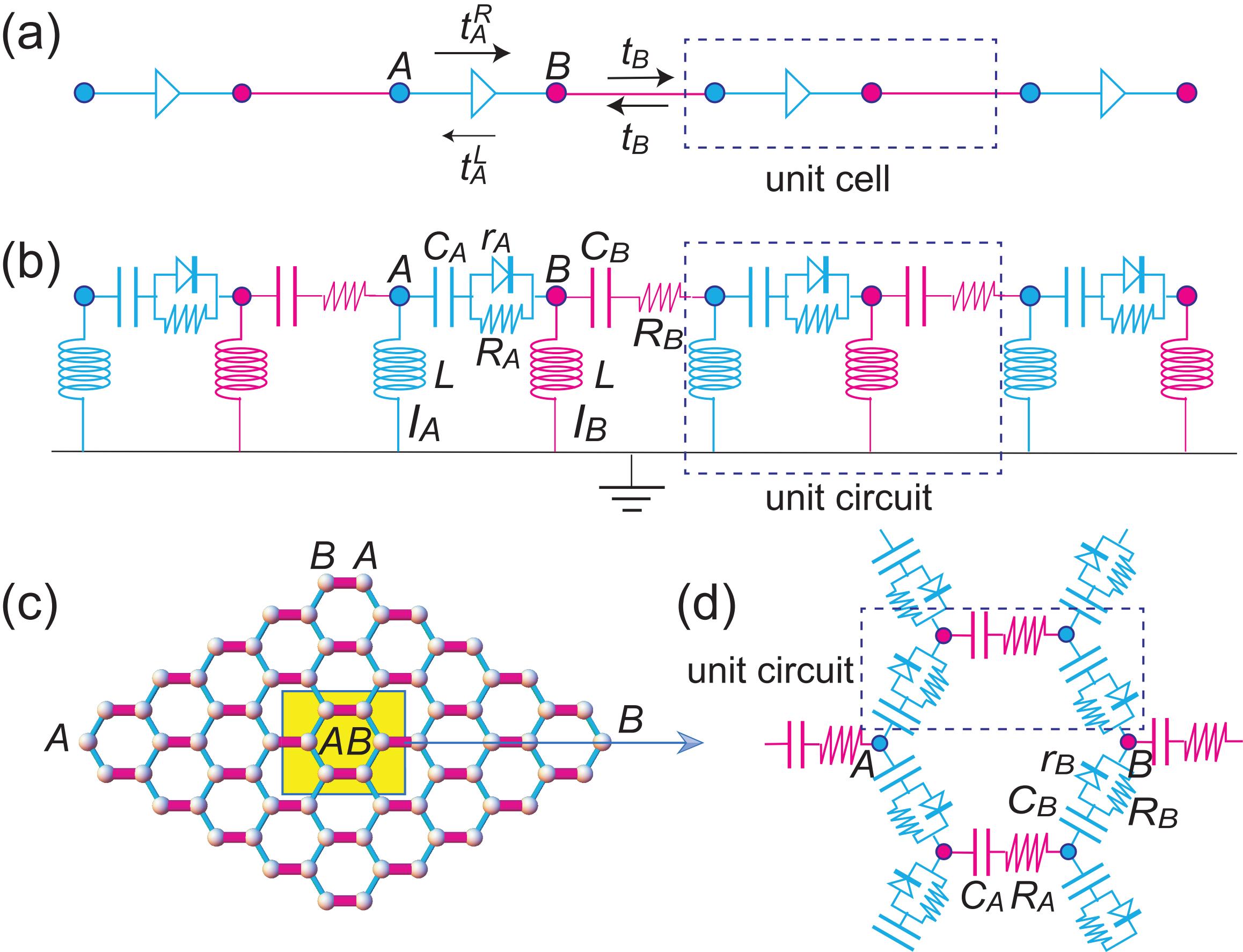}}
\caption{ (a) Illustration of the SSH model. One unit cell contains two
sites $A$ and $B$. Nonreciprocal links are shown by the symbol $\rhd$. (b)
Illustration of the SSH circuit. The reciprocal link is represented by a
condenser and a resistance connected in series. The nonreciprocal link is
obtained by replacing this resistance with a set of a diode and a resistance
connected in parallel. (c) Illustration of the anisotropic honeycomb model.
(d) Illustration of the electric-circuit realization. Each node is connected
to the ground via inductance $L$ as in (b). }
\label{FIG2}
\end{figure}

In this work, first we clarify the condition for the skin states to develop
in the non-Hermitian SSH model. The condition is found to be the
nonreciprocity of the hopping between the lattice sites. We then generalize
the analysis to higher dimensions. The Hermitian SSH model has been
generalized to higher dimensions such the anisotropic honeycomb and diamond
lattice models, where they are shown to be higher-order topological
insulators\cite{Science,Peng,Lang,Song,EzawaKagome,PhosHOTI}. We investigate
nonreciprocal versions of these models, and demonstrate the emergence of the
topological corner modes within the skin states: See Fig.\ref{FIG1}.

\begin{figure}[t]
\centerline{\includegraphics[width=0.48\textwidth]{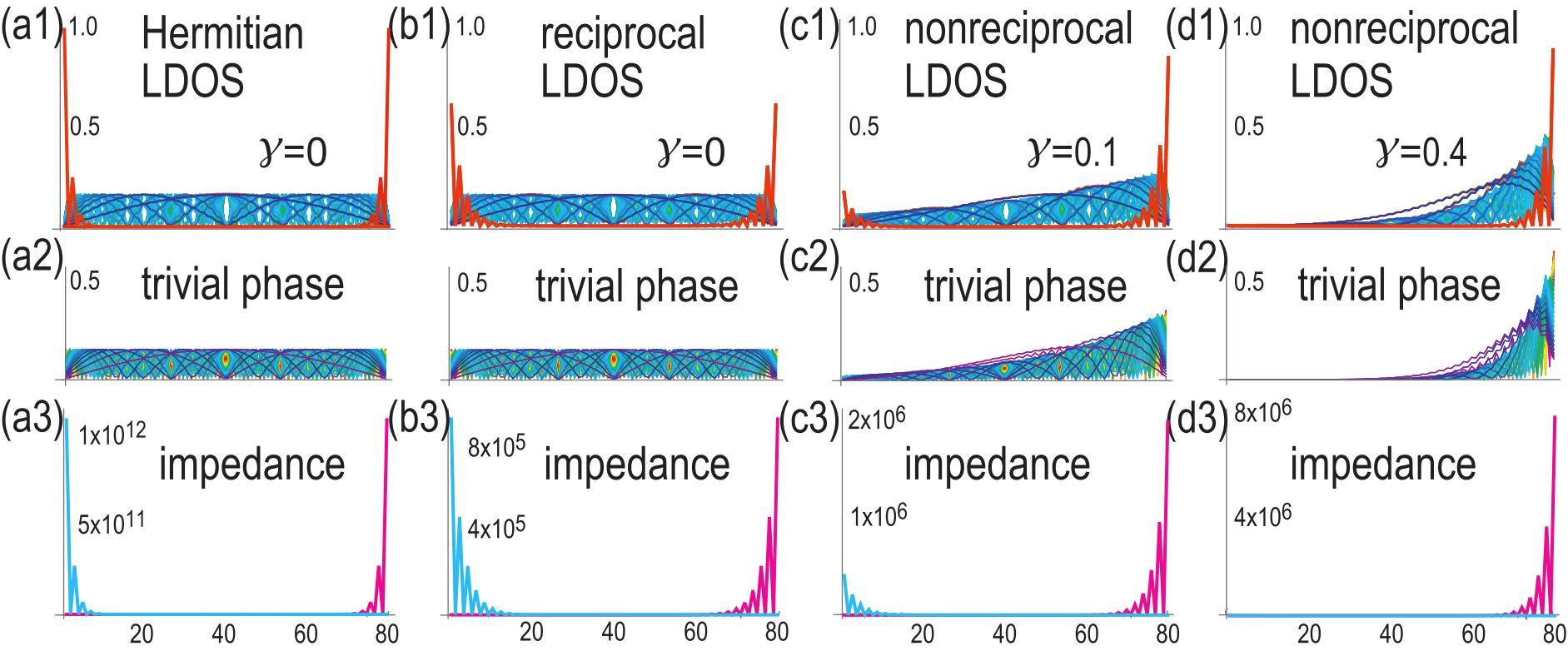}}
\caption{LDOS in the topological and trivial phases of the SSH model for
(a1)--(d1) and (a2)--(d2), respectively. The horizontal axis denotes the
lattice site number. Red curves represent the topological edge modes in
(a1)--(d1). (a1) LDOS of the Hermitian system, where the topological edge
modes are prominent at both edges. (b1) LDOS of the reciprocal non-Hermitian
system, which is quite similar to that of the Hermitian system. (c1)--(d1)
LDOS of the nonreciprocal non-Hermitian system, where the skin states are
formed. (a3)--(d3) Impedance (in unit of $\Omega $) in the corresponding SSH
circuits. We have taken $t_A=0.25$, $t_{B}=1$ and $\protect\gamma =0$ for
(a), $t_A=0.25+0.5i$, $t_{B}=1+0.5i$ and $\protect\gamma =0$ for (b), 
while $t_A=0.25+0.5i$, $t_{B}=1+0.5i$ and $\protect\gamma =0.1$ for (c) 
while $t_A=0.25+0.5i$, $t_{B}=1+0.5i$ and $\protect\gamma =0.4$ for (d). }
\label{FIG3}
\end{figure}

Electric circuits realize various topological phases\cite%
{ComPhys,TECNature,Garcia,Hel,Lu,EzawaTEC}. We show that $LCR$ circuits
with (without) diodes present a concrete playground to investigate
nonreciprocal (reciprocal) non-Hermitian topological physics (see Fig.\ref{FIG2}), 
where resistors naturally lead to non-Hermitian terms and diodes to
nonreciprocal terms. We focus on the chiral symmetric topological electric
circuits. When we analyze the SSH model, the anisotropic honeycomb and
diamond lattice models in $LC$ circuits, there are many impedance resonances
both in the topological and trivial phases. All of them are drastically
suppressed except for the topological zero-admittance modes due to the
effect of resistors in $LCR$ circuits. We then investigate their
nonreciprocal versions by introducing diodes, and conclude that the
emergence of topological corner modes in skin states is clearly detectable
by impedance peaks.

\textit{Non-Hermitian minimal two-band models:} The minimal model to
describe insulators is the two-band model. Indeed, one-band model cannot
have a line gap to generate insulators although it is
possible to have a point gap\cite{KawabataST}. The unit cell contains two
sites $A$ and $B$, and it is called a bipertite system. We investigate the
chiral symmetric model since it is known to have nontrivial topology
protected by the symmetry. The model is described by the $2\times 2$
Hamiltonian $H$, which is expanded in terms of the Pauli matrices, 
\begin{equation}
H\left( \mathbf{k}\right) =\left( 
\begin{array}{cc}
0 & h_{1}\left( \mathbf{k}\right) \\ 
h_{2}\left( \mathbf{k}\right) & 0%
\end{array}
\right) .  \label{BasicHamil}
\end{equation}
It has the chiral symmetry $\sigma _{z}$ satisfying $\left\{ H\left( \mathbf{%
k}\right) ,\sigma _{z}\right\} =0$, which assures the symmetric spectrum $E\leftrightarrow -E$. 
The diagonal terms are prohibited by the chiral
symmetry. Typical examples are the SSH model in one dimension, the
anisotropic honeycomb lattice in two dimensions, and the anisotropic diamond
lattice in three dimensions. See Figs.\ref{FIG1} and \ref{FIG2} for
illustrations of these lattices and also the corresponding electric
circuits. It is non-Hermitian when $h_{2}\left( \mathbf{k}\right) \neq
h_{1}^{\ast }\left( \mathbf{k}\right) $.

The quantum mechanical Hamiltonian describes hopping between two adjacent
sites. When the absolute value of the hopping toward one direction is equal
to the one toward the opposite direction, the system is said reciprocal and
otherwise nonreciprocal. It is intriguing that the hopping amplitude may be
complex. Indeed, complex hopping parameters appear naturally in an
electric-circuit realization of the non-Hermitian systems: See Eq.(\ref%
{HoppingCompl}). We have reciprocal and nonreciprocal non-Hermitian systems.

\textit{Non-Hermitian SSH models:} We start with the non-Hermitian
SSH model\cite{Yin,Yao,Kunst}, where $h_{1}\left( k\right)
=t_{A}^{L}+t_{B}e^{-ik}$ and $h_{2}\left( k\right) =t_{A}^{R}+t_{B}e^{ik}$. 
We illustrate the hopping parameters in Fig.\ref{FIG2}(a) and they
are complex in general. The system is reciprocal for $\left\vert
t_{A}^{L}\right\vert =\left\vert t_{A}^{R}\right\vert $, and nonreciprocal
otherwise. Let us set $t_{A}^{L}=t_{A}+\gamma /2$, $t_{A}^{R}=t_{A}-\gamma/2$%
, and call $\gamma$ the nonreciprocity.

The topological and trivial phases are distinguished by the emergence of
zero-energy edge modes for a finite chain. We define the local density
of states (LDOS) for the $n$-th right-eigen state $\left\vert
\psi _{n}^{\text{R}}\left( x\right) \right\rangle $\ by $\left\vert
\left\vert \psi _{n}^{\text{R}}\left( x\right) \right\rangle \right\vert ^{2}$. 
We show the LDOS for all $n$ with a choice of typical values of hopping parameters for a finite
chain in Fig.\ref{FIG3}. (i) The Hermitian SSH model is described by taking
a real parameter, $t_{A}\equiv t_{A}^{L}=t_{A}^{R}$. There are zero-energy
modes at both edges in the topological phase with $\left\vert
t_{A}\right\vert <\left\vert t_{B}\right\vert $ but none in the trivial
phase with $\left\vert t_{A}\right\vert >\left\vert t_{B}\right\vert $: See
Fig.\ref{FIG3}(a1)--(a2). (ii) We consider the reciprocal non-Hermitian
system by taking a complex value for $t_{A}=t_{A}^{L}=t_{A}^{R}$. As in Fig.%
\ref{FIG3}(b1), the LDOS is quite similar to the Hermitian SSH model though
the bulk energy becomes complex. The phase transition point is the same,
i.e., at $\left\vert t_{A}\right\vert =\left\vert t_{B}\right\vert $. (iii)
In Fig.\ref{FIG3}(c1)--(d1), we show the LDOS for the nonreciprocal
non-Hermitian system\cite{Yao} with $|t_{A}^{L}|\neq |t_{A}^{R}|$, which
demonstrates the formation of the skin states. By examining the zero-energy
edge mode, the system is topological for $\left\vert
t_{A}^{L}t_{A}^{R}\right\vert <\left\vert t_{B}\right\vert ^{2}$ and trivial
for $\left\vert t_{A}^{L}t_{A}^{R}\right\vert >\left\vert t_{B}\right\vert
^{2}$. Skin states are induced by the nonreciprocity both in the topological
and trivial phases. The zero-energy mode emerges only at one edge in the
topological phase. This is called the biorthogonal bulk-boundary
correspondence\cite{Yao,Kunst}.
We present some analytic formulas to understand the LDOS elsewhere\cite{SM-II}.

\begin{figure}[t]
\centerline{\includegraphics[width=0.49\textwidth]{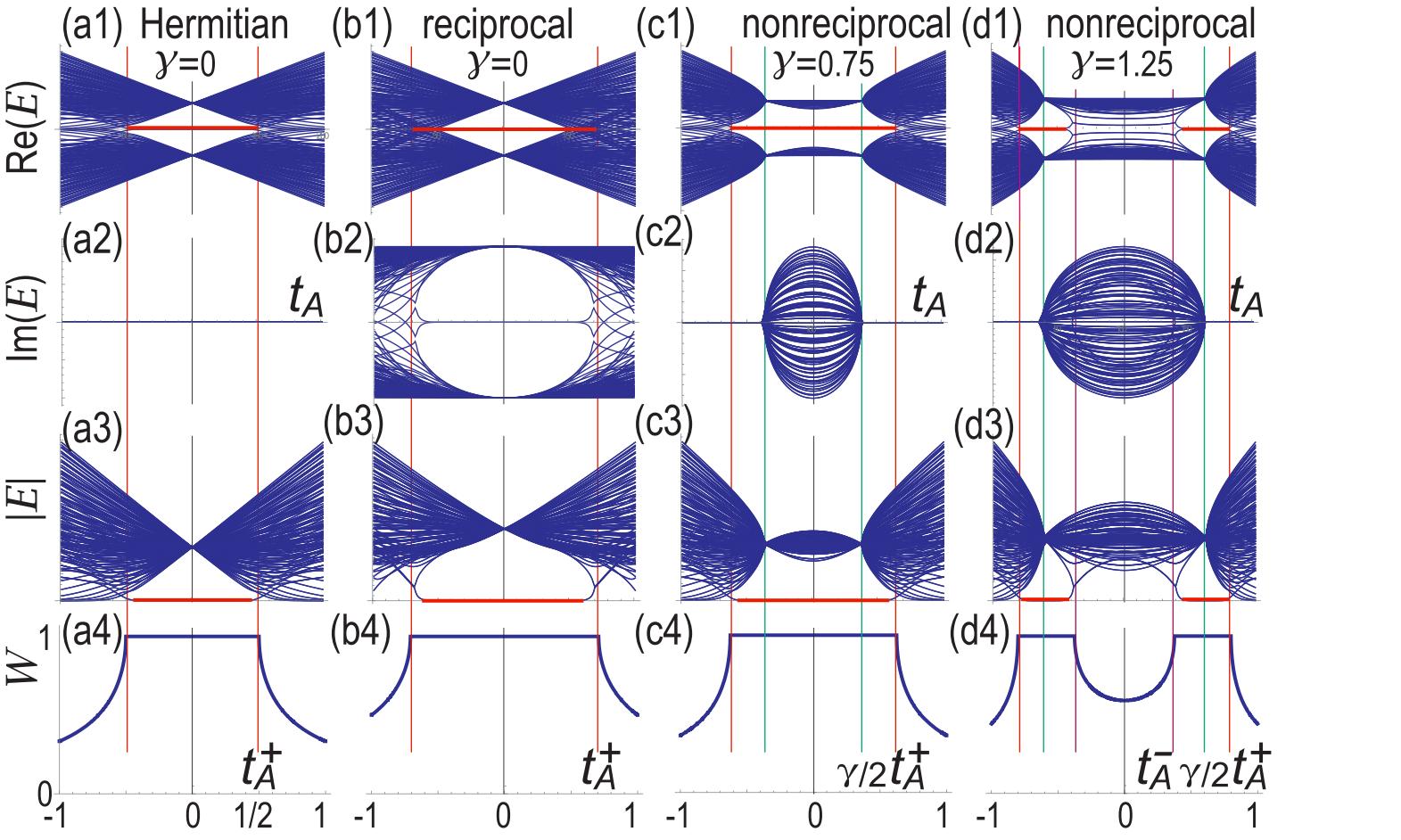}}
\caption{Energy spectra and topological numbers of the rhombus made of the
anisotropic honeycomb lattice. The horizontal axis is $t_{A}$, while the
vertical axis is the real part of the energy for (*1), the imaginary part
for (*2), the absolute value for (*3), and the topological number $W$ for
(*4). The horizontal red lines represent the topological corner modes, where
the system is a second-order topological insulator. (a*) for the Hermitian
model, (b*) for the reciprocal non-Hermitian model, (c*)--(d*) for the
nonreciprocal non-Hermitian model with increasing $\protect\gamma $. The
vertical lines represent the phase transition points at the value $t_{A}^{+}$
(magenta), $t_{A}^{-}$ (violet) and $\protect\gamma /2$ (green). We have
taken $t_{B}=1$ for (a), $t_{B}=1+i $ for (b), $t_{B}=1$ for (c), $t_{B}=1$
for (d).}
\label{FIG4}
\end{figure}

\textit{Non-Bloch winding numbers:} The non-Bloch topological number\cite%
{Yao,Yao2,Yang} describes the nonreciprocal non-Hermitian SSH model. Here,
we use the chiral index $\Gamma $\ as the non-Bloch topological number.
It is defined by\cite{SM-I}
\begin{equation}
\Gamma =\frac{1}{2i}\int_{-\pi }^{\pi }\frac{dk}{2\pi }\text{Tr}\left[
\sigma _{z}H\left( k+i\kappa \right) ^{-1}\partial _{k}H\left( k+i\kappa
\right) \right]   \label{ChiralIndex}
\end{equation}
in the two-band model (\ref{BasicHamil}), 
where $\kappa =-\log \sqrt{\left\vert t_{A}^{R}/t_{A}^{L}\right\vert }$. It
is zero, $\kappa =0$, for the reciprocal system, where this formula is
reduced to the usual chiral index. It is quantized as long as the chiral
symmetry is preserved. In addition, it cannot change its value as long as
the Hamiltonian is not singular. To see this, by substituting (\ref%
{BasicHamil}) to (\ref{ChiralIndex}), we obtain 
\begin{equation}
\Gamma =\frac{1}{2}\int_{-\pi }^{\pi }\frac{dk}{2\pi i}\partial _{k}\log
[h_{2}(k+i\kappa )/h_{1}(k+i\kappa )].
\end{equation}
Hence, the chiral index counts how many times the Hamiltonian winds the
origin. By evaluating it, we find that the system is topological ($\Gamma =1$
) for $\left\vert t_{A}^{L}t_{A}^{R}\right\vert <\left\vert t_{B}\right\vert
^{2}$ and trivial ($\Gamma =0$) for $\left\vert
t_{A}^{L}t_{A}^{R}\right\vert >\left\vert t_{B}\right\vert ^{2}$, as agrees
with the condition for the zero-energy edge modes to emerge in the skin
states.

A comment is in order. Although the chiral index (\ref{ChiralIndex}) has a
different expression from the non-Bloch topological number defined in Ref.\cite{Yao}, 
it is shown\cite{SM-I} that they are equivalent.

\textit{Non-Hermitian honeycomb lattices:} We proceed to investigate
non-Hermitian systems in two dimensions. A typical example is the
anisotropic honeycomb lattice model illustrated in Fig.\ref{FIG2}(c). The
nonreciprocal hopping is introduced as explained in Fig.\ref{FIG2}(d). The
Hamiltonian is given by the $2\times 2$ matrix (\ref{BasicHamil}) with $%
h_{1}=2t_{A}^{L}\cos (k_{y}/2)+t_{B}\exp (-ik_{x})$ and $h_{2}=2t_{A}^{R}%
\cos (k_{y}/2)+t_{B}\exp (ik_{x})$. The chiral index (\ref{ChiralIndex}) is
generalized to $D$ dimensions as 
\begin{equation}
\Gamma =\frac{1}{2i}\int \frac{d^{D}\mathbf{k}}{\left( 2\pi \right) ^{D}} 
\text{Tr}\left[ \sigma _{z}H\left( \mathbf{k}+i\mathbf{\kappa }\right)
^{-1}\partial _{k_{x}}H\left( \mathbf{k}+i\mathbf{\kappa }\right) \right] ,
\end{equation}
where the integration is performed over the Brillouin zone with $\mathbf{%
\kappa }=\left( \kappa ,\mathbf{0}\right) $. It is quantized when the system
is an insulator, while it changes its value continuously when the system is
metal. The system is topological for $\left\vert
t_{A}^{L}t_{A}^{R}/t_{B}^{2}\right\vert <1$. The topological phase
transition occurs from a topological insulator to a metal at $t_{A}^{+}=\pm 
\frac{1}{2}\sqrt{\left\vert t_{B}\right\vert ^{2}+\gamma ^{2}}$ for $%
\left\vert t_{B}\right\vert >\left\vert \gamma \right\vert $. On the other
hand, there are additional topological phase transitions at $t_{A}^{-}=\pm 
\frac{1}{2}\sqrt{-\left\vert t_{B}\right\vert ^{2}+\gamma ^{2}}$ for $%
\left\vert t_{B}\right\vert <\left\vert \gamma \right\vert $: See Fig.\ref{FIG4}. 
We give the derivation of $t_{A}^{\pm }$ elsewhere\cite{SM-III}.

\begin{figure}[t]
\centerline{\includegraphics[width=0.49\textwidth]{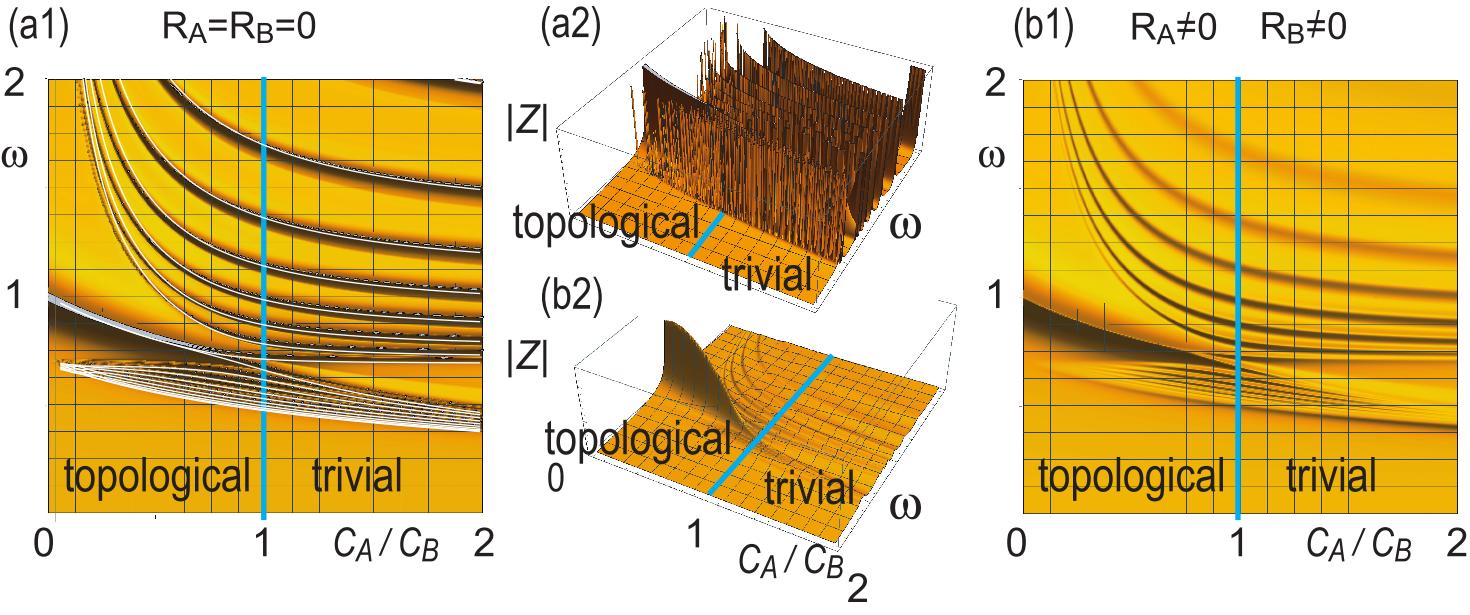}}
\caption{Impedance peaks are shown in the $(C_{A}/C_B)$-$\protect\omega $
plane. Phase transition occurs at $C_{A}=C_B$. (a) Many impedance peaks are
generated as implied by Eq.(\protect\ref{ResonFrequ}) both in the
topological and trivial phases of the Hermitian SSH model ($R_{A}=R_{B}=0$).
White curves represent the analytical result (\protect\ref{ResonFrequ}). (b)
All these peaks are suppressed drastically except for the topological peak
in the topological phase of the reciprocal non-Hermitian SSH model ($%
R_{A}\not=0,R_{B}\not=0$).}
\label{FIG5}
\end{figure}

We show the energy spectra and the topological numbers for various values of
hopping parameters for a rhombus made of the anisotropic honeycomb lattice
in Fig.\ref{FIG4}. (i) The Hermitian model is described by a real value for $%
t_{A}=t_{A}^{L}=t_{A}^{R}$. The Hermitian model produces a second-order
topological insulator in the parameter region $\left\vert
t_{A}/t_{B}\right\vert <1/2$. Namely, when we consider a nanoribbon, there
are no topological edge modes. On the other hand, topological corner modes
emerge at two corners in a rhombus as in Fig.\ref{FIG1}(a). These corner
modes are observed as zero-energy modes (depicted in red lines) in Fig.\ref%
{FIG4}(a). (ii) The reciprocal non-Hermitian system is constructed by taking
a complex value for $t_{A}=t_{A}^{L}=t_{A}^{R}$. The structure of the real
part of the energy spectrum and the topological charge are quite similar to
those of the Hermitian model, though the energy becomes complex as in Fig.%
\ref{FIG4}(b2). (iii) We consider the nonreciprocal non-Hermitian systems in
Fig.\ref{FIG4}(c) and (d). The zero-energy corner mode emerges at one of two
corners in a rhombus, as is found by calculating the LDOS: See Fig.\ref{FIG1}%
. These corner modes are observed as zero-energy modes (depicted in red
lines) in Fig.\ref{FIG4}(c) and (d).

In the similar way, we may analyze the skin states and the topological
corner mode in rhombohedron geometry of the diamond lattice. The Hamiltonian
is given by the $2\times 2$ matrix (\ref{BasicHamil}) with $%
h_{1}=t_{A}^{L}\left( 2\cos (k_{y}/2)+\exp \left( -ik_{z}\right)
\right)+t_{B}\exp (-ik_{x})$ and $h_{2}=t_{A}^{R}\left( 2\cos (k_{y}/2)+\exp
\left(ik_{z}\right) \right) +t_{B}\exp (ik_{x})$. The lattice structure is
illustrated in Fig.\ref{FIG1}, where the nonreciprocity is introduced just
as in Fig.\ref{FIG2}(d). The LDOS is also shown in Fig.\ref{FIG1}, which
demonstrates the formation of skin states and the topological corner mode.
It is a typical example of the nonreciprocal third-order topological
insulators in three dimensions\cite{SM-IV}.

\textit{Electric-circuit realization of non-Hermitian systems:} We consider
a class of electric circuits, where each node $a$ is connected to the ground
via inductance $L$: See Fig.\ref{FIG2}(b). Let $I_{a}$ be the current
between node $a$ and the ground via the inductance, $V_{a}$ be the voltage
at node $a$, $C_{ab}$ and $R_{ab}$ be the capacitance and the resistance
connected in series between nodes $a$ and $b$, respectively. We use diodes
to implement nonreciprocity in the circuit. We approximate a diode by a
linear resistance $r_{ab}$ for $a<b$ and the perfect nonreciprocity $%
r_{ab}=\infty$ for $b<a$. We set $C_{AB}=C_{A}$, $C_{BA}=C_{B}$, $%
R_{AB}^{R}=r_{A}R_{A}/(r_{A}+R_{A})$, $R_{AB}^{L}=R_{A}$, $%
R_{BA}^{R}=R_{BA}^{L}=R_{B}$ in a bipertite system.

The Kirchhoff's current law leads to the circuit Laplacian\cite%
{ComPhys,TECNature} $J_{ab}(\omega )$ with $\omega $ the frequency, 
\begin{equation}
J_{ab}\left( \omega \right) =i\omega \delta _{ab}[-\frac{1}{\omega ^{2}L}
+\sum_{c\neq a}H_{ac}(\omega )]-i\omega H_{ab}\left( \omega \right) ,
\end{equation}
where $H_{ab}\left( \omega \right) =C_{ab}/\left( 1+i\omega C_{ab}\bar{R}%
_{ab}\right) $. Here, $\bar{R}_{ab}=R_{ab}$ for $a>b$, $\bar{R}%
_{ab}=r_{ab}R_{ab}/(r_{ab}+R_{ab})$ for $a<b$, and $C_{aa}=0$ and $\bar{R}%
_{aa}=0$. An important observation is that $H_{ab}\left( \omega \right) $ is
identified as the tight-binding Hamiltonian in condensed-matter physics,
where the hopping parameters between adjacent sites $a$ and $b$ are given by 
\begin{align}
t_{A}^{L} &=C_{A}/\left( 1+i\omega C_{A}R_{A}\right) ,\quad
t_{B}=C_{B}/\left( 1+i\omega C_{B}R_{B}\right) ,  \notag \\
t_{A}^{R} &=C_{A}/[1+i\omega C_{A}/\left( 1/R_{A}+1/r_{A}\right) ].
\label{HoppingCompl}
\end{align}
The hopping parameters become complex.

\begin{figure}[t]
\centerline{\includegraphics[width=0.49\textwidth]{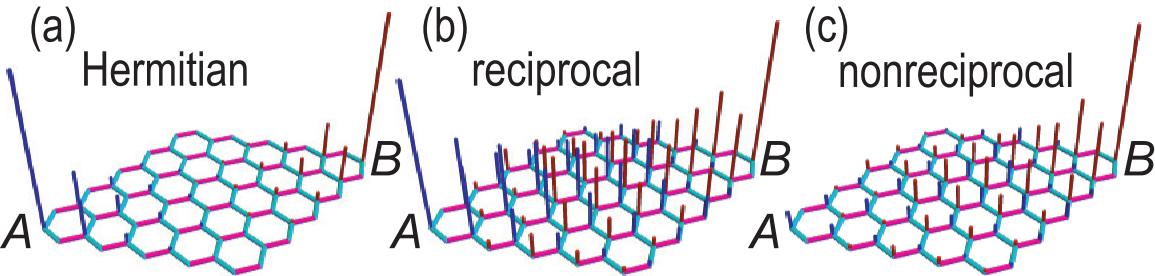}}
\caption{Spatial distribution of impedance in the topological phase of the
anisotropic honeycomb circuit. (a) In the Hermitian system, enhanced
topological peaks emerge at both corners. (b) In the reciprocal
non-Hermitian system topological peaks emerge at both corners, but they are
not so prominent. (c) In the nonreciprocal system, an enhanced topological
peak emerges only at corner $B$ corresponding to the topological skin corner
states around corner $B$ as in Fig.\protect\ref{FIG2}. }
\label{FIG6}
\end{figure}

\textit{Admittance spectrum and impedance peaks:} The admittance spectrum
consists of the eigenvalues of the circuit Laplacian\cite%
{ComPhys,TECNature,Garcia,Hel,Lu,EzawaTEC}. It is identical to the band
structure in condensed-matter physics. Thus, the topological edge or corner
modes correspond to the zero-admittance modes.

A measurable quantity of electric circuits is the impedance, which is given
by\cite{Hel} $G_{ab}=V_{a}/I_{b}$, where $G$ is the green function defined
by the inverse of the Laplacian $J$, $G\equiv J^{-1}$. It diverges at the
frequency satisfying $J=0$. Therefore, it is possible to detect the
topological zero-admittance modes by the divergence of the impedance.

Let us first search for zero-admittance modes in the $LC$ circuit. After the
diagonalization, the circuit Laplacian reads 
\begin{equation}
J_{n}\left( \omega \right) =i\omega \lbrack -(\omega
^{2}L)^{-1}+\sum_{\alpha =A,B}n_{\alpha }C_{\alpha }]-i\omega \varepsilon
_{n}\left( \omega \right) ,
\end{equation}
where $n_{\alpha }$ is the number of the nodes adjacent to node $\alpha $,
and $\varepsilon _{n}$ is the eigenvalue of the circuit Laplacian. The
impedance diverges at the resonance frequencies 
\begin{equation}
\omega _{\text{R}}(\varepsilon _{n})=\sqrt{(-\varepsilon _{n}+\sum_{\alpha
}n_{\alpha }C_{\alpha })/L},  \label{ResonFrequ}
\end{equation}
which is the solution of $J_{n}\left( \omega \right) =0$. Hence there are
many impedance peaks indexed by $n$ both in the topological and trivial
phases as in Fig.\ref{FIG5}(a1)--(a2) for the instance of the SSH model,
among which the topological impedance peak is given by the zero-admittance
mode ($\varepsilon _{0}=0$) in the topological phase. However, when we
introduce resistors, since $\varepsilon _{n}$ becomes complex except for the
zero-admittance mode ($\varepsilon _{0}=0$), all resonance peaks are
drastically suppressed except for the topological peak as in Fig.\ref{FIG5}%
(b1)--(b2) for the instance of the SSH model. This phenomenon occurs in any
dimensions, since the resonance frequency (\ref{ResonFrequ}) is valid in any
dimensions.

We first present calculate the impedance at each node in the SSH model. We
show a space distribution of the point impedance in the topological phase in
Fig.\ref{FIG3}(a3)--(d3). We see how the topological impedance peak develops
in the skin states as the nonreciprocity $\gamma$ increases.

We next calculate the impedance at each node in rhombus (rhombohedron)
geometry of the anisotropic honeycomb (diamond) lattice, where the
second-order (third-order) topological phase is realized and the topological
corner mode emerges. We show a space distribution of the point impedance in
the topological phase in Fig.\ref{FIG6}(a), (b) and (c) for the Hermitian,
reciprocal non-Hermitian and nonreciprocal non-Hermitian honeycomb systems,
respectively. In the reciprocal system, impedance peaks emerge at corners $A$
and $B$. On the other hand, in the nonreciprocal system, an impedance peak
emerges only at corner $B$ corresponding to the topological corner modes in
Fig.\ref{FIG1}.

\textit{Discussion:} We have studied a non-Hermitian extension of the
higher-order topological phases and proposed to realize them by
electric circuits. Our results show that various non-Hermitian systems will
be implimented in electric circuits.

After submission of this work, we find closely related works\cite{RefA,RefB,RefC} on
the non-Hermitian extensions of the higher-order topological phases.

The author is very much grateful to N. Nagaosa for helpful discussions on
the subject. This work is supported by the Grants-in-Aid for Scientific
Research from MEXT KAKENHI (Grants No. JP17K05490, No. JP15H05854 and No.
JP18H03676). This work is also supported by CREST, JST (JPMJCR16F1 and
JPMJCR1874).

%%%%%%%%%%%%%%%
\clearpage\newpage
\onecolumngrid
\def\theequation{S\arabic{equation}}
\def\thefigure{S\arabic{figure}}
\def\thesubsection{S\arabic{subsection}}
\setcounter{figure}{0}
\setcounter{equation}{0}

\centerline{\textbf{\Large Supplemental Material}}
%%%%%%%%%%%%%%%

\section{Non-Bloch topological number}

We summarize and compare several possible definitions of the topological
number in the chiral symmetric $2\times 2$ non-Hermitian Hamiltonian
systems. In particular, we prove the equivalence between the definition (2)
in the main text and the definition used in the original work\cite{Yao}.
Although the generalization to an arbitrary dimensional space is
straightforward just as in Eq.(3), we explicitly discuss the one-dimensional
models.

\subsection{Biorthogonal basis}

The Hamiltonian we investigate is of the following form,%
\begin{equation}
H=\left( 
\begin{array}{cc}
0 & h_{1} \\ 
h_{2} & 0%
\end{array}%
\right) .  \label{HamilA}
\end{equation}%
It is necessary to take the biorthogaonal basis in the
non-Hermitian theory, by introducing the right eigen-function $\left\vert
\psi _{\pm }^{\text{R}}\right\rangle $ satisfying $H\left\vert \psi _{\pm }^{%
\text{R}}\right\rangle =\pm \varepsilon ^{\text{R}}\left\vert \psi _{\pm }^{%
\text{R}}\right\rangle $ and the left eigen-function $\left\vert \psi _{\pm
}^{\text{L}}\right\rangle $ satisfying $H^{\dagger }\left\vert \psi _{\pm }^{%
\text{L}}\right\rangle =\pm \varepsilon ^{\text{L}}\left\vert \psi _{\pm }^{%
\text{L}}\right\rangle $. They are given by%
\begin{equation}
\left\vert \psi _{\pm }^{\text{R}}\right\rangle =\frac{1}{\sqrt{2}}\left\{ 
\sqrt{\frac{h_{1}}{h_{2}}},\pm 1\right\} ,\quad \left\langle \psi _{\pm }^{%
\text{L}}\right\vert =\frac{1}{\sqrt{2}}\left\{ \sqrt{\frac{h_{2}}{h_{1}}}%
,\pm 1\right\} ,
\end{equation}%
with the eigen energy%
\begin{equation}
\varepsilon ^{\text{L}}=\left( \varepsilon ^{\text{R}}\right) ^{\ast }=\sqrt{%
h_{1}h_{2}}.
\end{equation}%
They satisfy the biorthogonal condition%
\begin{equation}
\left\langle \psi _{\pm }^{\text{L}}\left\vert \psi _{\pm }^{\text{R}%
}\right\rangle \right. =1,\qquad \left\langle \psi _{\pm }^{\text{L}%
}\left\vert \psi _{\mp }^{\text{R}}\right\rangle \right. =0.
\end{equation}%
They are reduced to the ordinary eigen-function and eigen energy in the
Hermitian limit.

\subsection{Chiral index and Non-Hermitian winding number}

We define the chiral index $\Gamma $ for the Hermitian system (\ref{HamilA})
by the formula%
\begin{equation}
\Gamma =\frac{1}{2i}\int_{-\pi }^{\pi }\frac{dk}{2\pi }\text{Tr}\left[
\sigma _{z}H^{-1}\partial _{k}H\right] =\frac{1}{2}(W_{2}-W_{1}),
\end{equation}%
where%
\begin{equation}
W_{\alpha }=\int_{-\pi }^{\pi }\frac{dk}{2\pi i}\partial _{k}\log h_{\alpha
}(k).
\end{equation}%
It is a winding number, which is quantized even in the nonreciprocal
non-Hermitian system. This is a good candidate for the topological
number. However, it does not give the correct phase transition point at
which the topological zero-energy edge mode emerges in the skin state. The
correct one is given with the introduction of the complex momentum\cite{Yao}
as%
\begin{equation}
W_{\alpha }=\int_{-\pi }^{\pi }\frac{dk}{2\pi i}\partial _{k}\log h_{\alpha
}(k+i\kappa ),  \label{WindiNon}
\end{equation}%
where $\kappa =-\log \sqrt{\left\vert t_{A}^{R}/t_{A}^{L}\right\vert }$. It
is understood that the argument of $h_{\alpha }$ is $k+i\kappa $ in what
follows.

\subsection{Non-Bloch winding number}

The original definition of the non-Bloch topological number\cite{Yao} reads
as follows. First, the $Q$ matrix is defined by%
\begin{equation}
Q\equiv \left\vert \tilde{\psi}_{-}^{\text{R}}\right\rangle \left\langle 
\tilde{\psi}_{-}^{\text{L}}\right\vert -\left\vert \psi _{-}^{\text{R}%
}\right\rangle \left\langle \psi _{-}^{\text{L}}\right\vert =-\left( 
\begin{array}{cc}
0 & \sqrt{h_{1}/h_{2}} \\ 
\sqrt{h_{2}/h_{1}} & 0%
\end{array}%
\right) \equiv \left( 
\begin{array}{cc}
0 & q \\ 
q^{-1} & 0%
\end{array}%
\right) ,
\end{equation}%
with%
\begin{equation}
q=-\sqrt{h_{1}/h_{2}},
\end{equation}%
and 
\begin{equation}
\left\vert \tilde{\psi}_{-}^{\text{R}}\right\rangle =\sigma _{z}\left\vert
\psi _{-}^{\text{R}}\right\rangle ,\quad \left\vert \tilde{\psi}_{-}^{\text{L%
}}\right\rangle =\sigma _{z}\left\vert \psi _{-}^{\text{L}}\right\rangle .
\end{equation}%
Then, the winding number is defined by%
\begin{equation}
W=\int_{-\pi }^{\pi }\frac{dk}{2\pi i}\partial _{k}\log q=\int_{-\pi }^{\pi }%
\frac{dk}{2\pi i}\frac{\partial _{k}\log h_{2}-\partial _{k}\log h_{1}}{2}=%
\frac{W_{2}-W_{1}}{2}.
\end{equation}%
It agrees with the chiral index $\Gamma $.

\subsection{Non-Hermitian Wannier center}

The Wannier center is the polarization vector in general, and has been
proposed to be the topological number in the chiral symmetric $2\times 2$
Hermitian Hamiltonian. Its generalization to the non-Hermitian system is
straightforward. It is given by\cite{Lieu}%
\begin{equation}
p=\frac{1}{2\pi }\int Adk=\frac{1}{2\pi i}\int \left\langle \psi ^{\text{L}%
}\right\vert \partial _{k}\left\vert \psi ^{\text{R}}\right\rangle dk,
\end{equation}%
where $A=-i\left\langle \psi ^{\text{L}}\right\vert \partial _{k}\left\vert
\psi ^{\text{R}}\right\rangle $ is the non-Hermitian Berry curvature. It is
identical to the winding number in the SSH model since%
\begin{equation}
p=\frac{1}{2\pi i}\int \frac{\partial _{k}\log h_{1}-\partial _{k}\log h_{2}%
}{4}dk=\frac{W_{1}-W_{2}}{4}=-\frac{\Gamma }{2}.
\end{equation}%
The non-Hermitian polarization is half quantized for the topological phase.

\subsection{Winding number of energy}

Since the energy is a complex number in the non-Hermitian SSH model, we may
define a topological charge as\cite{Fu,UedaPRX}%
\begin{equation}
W_{E}=\frac{1}{2\pi }\int \partial _{k}\log Edk=\frac{1}{2\pi }\int \partial
_{k}\log \sqrt{h_{1}h_{2}}dk=\frac{1}{4\pi }\int \left[ \partial _{k}\log
h_{1}+\partial _{k}\log h_{2}\right] dk=\frac{W_{1}+W_{2}}{2}.
\end{equation}%
It is zero in the SSH model. Indeed, it follows that $W_{1}=-W_{2}$ in the
SSH model with the explicit use of the functions $h_{1}$ and $h_{2}$.

\subsection{Winding number of the Hamiltonian}

We also define the winding number for the Hamiltonian\cite{UedaPRX}

\begin{equation}
W_{H}=\int_{-\pi }^{\pi }\frac{dk}{2\pi i}\text{Tr}\left[ H^{-1}\partial
_{k}H\right] =W_{1}+W_{2}=2W_{E}.
\end{equation}%
It is also zero in the SSH model.

\section{Non-Hermitian SSH model}

The non-Hermitian SSH model is given by (\ref{HamilA}) with%
\begin{equation}
h_{1}\left( k\right) =t_{A}^{L}+t_{B}e^{-ik},\qquad h_{2}\left( k\right)
=t_{A}^{R}+t_{B}e^{ik}.
\end{equation}%
We thus find%
\begin{equation}
h_{1}(k+i\kappa )=t_{A}^{L}+t_{B}\sqrt{\frac{t_{A}^{L}}{t_{A}^{R}}}%
e^{-ik},\qquad h_{2}(k+i\kappa )=t_{A}^{R}+t_{B}\sqrt{\frac{t_{A}^{R}}{%
t_{A}^{L}}}e^{ik}.
\end{equation}%
The winding numbers (\ref{WindiNon}) are calculated as 
\begin{equation}
W_{1}=\left\{ 
\begin{array}{ccc}
-1 & \text{for} & \left\vert \frac{\sqrt{t_{A}^{L}t_{A}^{R}}}{t_{B}}%
\right\vert <1 \\ 
0 & \text{for} & \left\vert \frac{\sqrt{t_{A}^{L}t_{A}^{R}}}{t_{B}}%
\right\vert >1%
\end{array}%
\right. ,\qquad W_{2}=\left\{ 
\begin{array}{ccc}
1 & \text{for} & \left\vert \frac{\sqrt{t_{A}^{L}t_{A}^{R}}}{t_{B}}%
\right\vert <1 \\ 
0 & \text{for} & \left\vert \frac{\sqrt{t_{A}^{L}t_{A}^{R}}}{t_{B}}%
\right\vert >1%
\end{array}%
\right. ,
\end{equation}%
where we find the relation $W_{2}=-W_{1}$, and hence there is only one
winding number $W=\left( W_{2}-W_{1}\right) /2$ in the SSH model.

\subsection{Finite chain}

We review an analytic form of the eigen-functions at the zero-energy
state in the non-Hermitian SSH model\cite{Kunst}. We label the eigen-function at the outer most node
as $\psi _{1}$, and that of the node next to it as $\psi _{2}$, and as so
on. Namely, we write the eigen function as $\psi =\left\{ \psi _{1},\psi
_{2},\cdots ,\psi _{2N}\right\} $ if there are 2$N$ nodes across the chain.
The eigen equation is given by%
\begin{equation}
H\psi =\left( 
\begin{array}{cccccc}
0 & t_{A}^{L} & 0 & 0 & 0 & \cdots \\ 
t_{A}^{R} & 0 & t_{B}^{L} & 0 & 0 & \cdots \\ 
0 & t_{B}^{R} & 0 & t_{A}^{L} & 0 & \cdots \\ 
0 & 0 & t_{A}^{R} & 0 & t_{B}^{L} & \cdots \\ 
0 & 0 & 0 & t_{B}^{R} & 0 & \cdots \\ 
\cdots & \cdots & \cdots & \cdots & \cdots & \cdots%
\end{array}%
\right) \left( 
\begin{array}{c}
\psi _{1} \\ 
\psi _{2} \\ 
\psi _{3} \\ 
\psi _{4} \\ 
\psi _{5} \\ 
\cdots%
\end{array}%
\right) =E\left( 
\begin{array}{c}
\psi _{1} \\ 
\psi _{2} \\ 
\psi _{3} \\ 
\psi _{4} \\ 
\psi _{5} \\ 
\cdots%
\end{array}%
\right) .
\end{equation}%
The eigenvalue problem $H\psi =0$ is explicitly given by%
\begin{eqnarray}
t_{A}^{L}\psi _{2} &=&0,  \notag \\
t_{A}^{R}\psi _{1}+t_{B}^{L}\psi _{3} &=&0,  \notag \\
t_{B}^{R}\psi _{2}+t_{A}^{L}\psi _{4} &=&0, \\
t_{A}^{R}\psi _{3}+t_{B}^{L}\psi _{5} &=&0,  \notag \\
t_{B}^{R}\psi _{4}+t_{A}^{L}\psi _{6} &=&0,  \notag \\
&&\cdots  \notag
\end{eqnarray}%
By solving them recursively from the outer most cite, we obtain the relation 
\begin{equation}
\psi _{2n+1}=\left( -\frac{t_{A}^{R}}{t_{B}^{L}}\right) ^{n}\psi _{1},
\label{p1}
\end{equation}%
and the analytic form of the eigen mode for odd cite $n$. In the same way,
we obtain%
\begin{equation}
\psi _{2n}=\left( -\frac{t_{A}^{L}}{t_{B}^{R}}\right) ^{2N-n}\psi _{2N}.
\label{p2}
\end{equation}%
These solutions well reproduce numerical results of the zero-energy states
shown in Fig.3 of the main text.

\subsection{Impedance}

The impedance is given by the Green funcition, which is explicitly written
as 
\begin{equation}
G=H^{-1}=\left( 
\begin{array}{cccccccc}
0 & \frac{1}{t_{A}^{R}} & 0 & \frac{1}{t_{A}^{R}}\left( -\frac{t_{B}^{L}}{%
t_{A}^{R}}\right) & 0 & \frac{1}{t_{A}^{R}}\left( -\frac{t_{B}^{L}}{t_{A}^{R}%
}\right) ^{2} & 0 & \cdots \\ 
\frac{1}{t_{A}^{L}} & 0 & 0 & 0 & 0 & 0 & 0 & \cdots \\ 
0 & 0 & 0 & \frac{1}{t_{A}^{R}} & 0 & \frac{1}{t_{A}^{R}}\left( -\frac{%
t_{B}^{L}}{t_{A}^{R}}\right) & 0 & \cdots \\ 
\frac{1}{t_{A}^{L}}\left( -\frac{t_{B}^{R}}{t_{A}^{L}}\right) & 0 & \frac{1}{%
t_{A}^{L}} & 0 & 0 & 0 & 0 & \cdots \\ 
0 & 0 & 0 & 0 & 0 & \frac{1}{t_{A}^{R}} & 0 & \cdots \\ 
\frac{1}{t_{A}^{L}}\left( -\frac{t_{B}^{R}}{t_{A}^{L}}\right) ^{2} & 0 & 
\frac{1}{t_{A}^{L}}\left( -\frac{t_{B}^{R}}{t_{A}^{L}}\right) & 0 & \frac{1}{%
t_{A}^{L}} & 0 & 0 & \cdots \\ 
0 & 0 & 0 & 0 & 0 & 0 & 0 & \cdots \\ 
\cdots & \cdots & \cdots & \cdots & \cdots & \cdots & \cdots & \cdots%
\end{array}%
\right)
\end{equation}%
for the SSH model. Each component corresponds to the LDOS of the zero-energy
edge states (\ref{p1}) and (\ref{p2}). They well reproduce impedance peaks
shown in Fig.3 of the main text.

\section{Non-Hermitian honeycomb lattice}

The anisotropic honeycomb lattice is known to have a second-order
topological insulator phase for $\left\vert t_{A}/t_{B}\right\vert <1/2$ in
the Hermitian model\cite{PhosHOTI}. The non-Hermitian model is given by (\ref%
{HamilA}) with

\begin{equation}
h_{1}=2t_{A}^{L}\cos (k_{y}/2)+t_{B}\exp (-ik_{x}),\qquad
h_{2}=2t_{A}^{R}\cos (k_{y}/2)+t_{B}\exp (ik_{x}).
\end{equation}%
The topological charge in higher dimensions is given by $W=\left(
W_{2}-W_{1}\right) /2$ with (3) in the main text. We find $W_{2}=-W_{1}$ as
in the SSH model.

The system is topological for 
\begin{equation}
\sqrt{\left\vert \frac{t_{A}^{L}t_{A}^{R}}{t_{B}^{2}}\right\vert }<\frac{1}{2%
},
\end{equation}%
where we have used the relation%
\begin{equation}
2\left\vert \cos (\sqrt{3}k_{x}/2)\right\vert \sqrt{\left\vert \frac{%
t_{A}^{L}t_{A}^{R}}{t_{B}^{2}}\right\vert }<2\sqrt{\left\vert \frac{%
t_{A}^{L}t_{A}^{R}}{t_{B}^{2}}\right\vert }.
\end{equation}%
The topological phase transitions occur at%
\begin{equation}
\left\vert \frac{t_{A}^{L}t_{A}^{R}}{t_{B}^{2}}\right\vert =\frac{1}{2},
\end{equation}%
which are equal to%
\begin{equation}
t_{A}^{+}=\pm \frac{1}{2}\sqrt{\left\vert t_{B}\right\vert ^{2}+\gamma ^{2}}.
\end{equation}%
For $\gamma >\left\vert t_{B}\right\vert $, there are additional phase
transitions at%
\begin{equation}
t_{A}^{-}=\pm \frac{1}{2}\sqrt{-\left\vert t_{B}\right\vert ^{2}+\gamma ^{2}}%
.
\end{equation}%
These phase transition points are found in Fig.4(c) and (d) in the main text.

The main results on the anisotropic honeycomb lattice are the LDOS given in
Fig.1, the energy spectrum and the topological number given in Fig.4 and the
impedance peaks given in Fig.6 in the main text. We have also shown how the
skin states evolves in Fig.1 as the nonreciprocity is increased.

\begin{figure}[t]
\centerline{\includegraphics[width=0.7\textwidth]{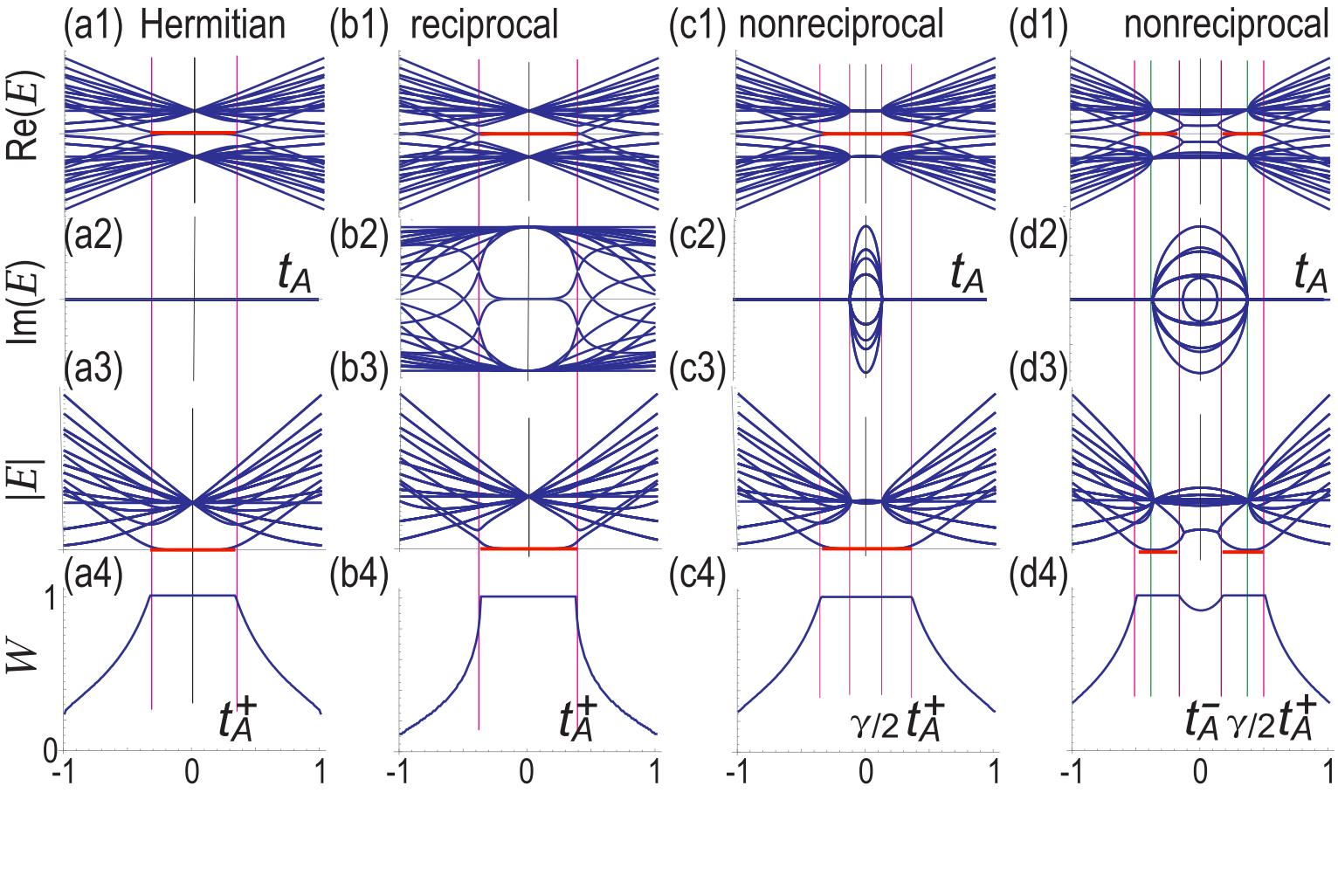}}
\caption{Energy spectra and topological numbers of the rhombohedron made of
the anisotropic diamond lattice. The horizontal axis is $t_{A}$, while the
vertical axis is the real part of the energy for (*1), the imaginary part
for (*2), the absolute value for (*3), and the topological number $W$ for
(*4). The horizontal red lines represent the topological corner modes, where
the system is a third-order topological insulator. (a*) for the Hermitian
model, (b*) for the reciprocal non-Hermitian model, (c*) for the
nonreciprocal non-Hermitian model with increasing $\protect\gamma $. The
vertical lines represent the phase transition points at the value $t_{A}^{+}$
(magenta), $t_{A}^{-}$ (violet) and $\protect\gamma /2$ (green). We have
taken $\protect\gamma =0$ and $t_{B}=1$ for (a), $\protect\gamma =0$ and $%
t_{B}=1+i $ for (b), $\protect\gamma =0.25$ and $t_{B}=1$ for (c), $\protect%
\gamma =0.75$ and $t_{B}=1$ for (d). }
\label{FigDiamondEne}
\end{figure}

\section{Non-Hermitian diamond lattice}

The anisotropic diamond lattice is known to have a third-order topological
insulator phase for $\left\vert t_{A}/t_{B}\right\vert <1/3$ in the
Hermitian model\cite{PhosHOTI}. The non-Hermitian model is given by (\ref%
{HamilA}) with%
\begin{equation}
h_{1}=t_{A}^{L}\left( 2\cos (k_{y}/2)+\exp \left( -ik_{z}\right) \right)
+t_{B}\exp (-ik_{x}),\qquad h_{2}=t_{A}^{R}\left( 2\cos (k_{y}/2)+\exp
\left( ik_{z}\right) \right) +t_{B}\exp (ik_{x}).
\end{equation}%
The topological charge is given by $W=\left( W_{2}-W_{1}\right) /2$ with (%
\ref{WindiNon}). We find $W_{2}=-W_{1}$ as in the SSH model. The system is
topological for 
\begin{equation}
\sqrt{\left\vert \frac{t_{A}^{L}t_{A}^{R}}{t_{B}^{2}}\right\vert }<\frac{1}{3%
},
\end{equation}%
where we have used the relation%
\begin{equation}
\left\vert 2\cos (\sqrt{3}k_{x}/2)+\exp (i\frac{3k_{y}}{2})\right\vert \sqrt{%
\left\vert \frac{t_{A}^{L}t_{A}^{R}}{t_{B}^{2}}\right\vert }<3\sqrt{%
\left\vert \frac{t_{A}^{L}t_{A}^{R}}{t_{B}^{2}}\right\vert }.
\end{equation}%
The topological phase transitions occur at%
\begin{equation}
\left\vert \frac{t_{A}^{L}t_{A}^{R}}{t_{B}^{2}}\right\vert =\frac{1}{3},
\end{equation}%
which are equal to%
\begin{equation}
t_{A}^{+}=\pm \frac{1}{6}\sqrt{4\left\vert t_{B}\right\vert ^{2}+9\gamma ^{2}%
}.
\end{equation}%
For $\gamma >\frac{2}{3}\left\vert t_{B}\right\vert $, there are additional
phase transitions at%
\begin{equation}
t_{A}^{-}=\pm \frac{1}{6}\sqrt{-4\left\vert t_{B}\right\vert ^{2}+9\gamma
^{2}}.
\end{equation}%
These phase transition points are found in Fig.\ref{FigDiamondEne}(c) and
(d).

The main results on the anisotropic diamond lattice are the LDOS given in
Fig.1 in the main text, the energy spectrum and the topological number given
in Fig.\ref{FigDiamondEne}. We have also shown how the skin states evolves
in Fig.1 as the nonreciprocity is increased.

\end{thebibliography}

\end{document}